\documentclass[12pt]{article}

\usepackage{psfig}
\usepackage{epsf}

\newcommand{\be}{\begin{eqnarray}}
\newcommand{\ee}{\end{eqnarray}}
\newcommand{\non}{\nonumber\\}
\newcommand{\mbf}[1]{{\mbox{\boldmath{$#1$}}}}
\newcommand{\colvec}[1]{{\begin{array}{c} #1\end{array}}}
\newcommand{\equ}[1]{Eq.~(\ref{#1})}
\newcommand{\hatD}{{\hat{D}}}
\newcommand{\ave}[1]{\left\langle #1 \right\rangle}
\newcommand{\bbox}[1]{\hbox{\boldmath{$#1$}}}
\newcommand{\mbfs}{{\mbf{s}}}
\newcommand{\bfl}{{\bf l}} 
\newcommand{\bra}[1]{\left\langle #1 \right|}
\newcommand{\ket}[1]{\left| #1 \right\rangle}
\newcommand{\bfQ}{{\bf Q}} 
\newcommand{\mbfN}{{\mbf{N}}}
\newcommand{\mbfR}{{\mbf{R}}}
\newcommand{\bfe}{{\bf e}} 

\title{
\vskip 0.1in
Random walks of partons in $SU(N_c)$ and classical representations of
color charges in QCD at small $x$}
\author{
Sangyong Jeon\\
{\small \it Physics Department, McGill University,Montreal, QC H3A-2T8,
Canada}\\
{\small \it  and RIKEN-BNL Research Center, Brookhaven National Laboratory,
Upton, NY 11973}\\
\\
Raju Venugopalan\\
{\small \it Physics Department,
Brookhaven National Laboratory, Upton, NY 11973.}\\
}

\begin{document}

\maketitle

\begin{center}
{\bf Abstract}

\vspace{0.2cm}

\begin{minipage}{0.9\textwidth}
The effective action for wee partons in large nuclei includes a sum
over static color sources distributed in a wide range of
representations of the $SU(N_c)$ color group. The problem can be
formulated as a random walk of partons in the $N_c-1$ dimensional
space spanned by the Casimirs of $SU(N_c)$. For a large number of
sources, $k \gg 1$, we show explicitly that the most likely
representation is a classical representation of order O($\sqrt{k}$).
The quantum sum over representations is well approximated by a path
integral over classical sources with an exponential weight whose
argument is the quadratic Casimir operator of the group. The
contributions of the higher $N_c-2$ Casimir operators are suppressed
by powers of $k$. Other applications of the techniques developed here
are discussed briefly.
\end{minipage}

\end{center}

\noindent

\vfill \eject

\baselineskip=22pt plus 1pt minus 1pt
\parindent=25pt

\vskip 0.1in

\section{Introduction}
\vskip 0.1in

In QCD, there are a number of physical situations where higher
dimensional representations of the gauge group are relevant. In
particular, in perturbative QCD, because of the ubiquitous nature of
bremsstrahlung processes, one often encounters situations where
partons at one energy scale interact simultaneously with several
partons at a different energy scale~\cite{StermanTASI95}.  The former
may therefore couple to color charges in a wide range of
representations of the color group. If one can argue that higher
dimensional representations are the most likely representations, it
may be possible, by the correspondence principle, to treat these as
classical representations. Classical color charges are considerably
simpler to treat than their quantum counterparts.

In this paper, we will address the following formal questions which
are relevant to addressing this issue quantitatively.  Given $k$
non-interacting quarks in the fundamental $SU(N_c)$ representation,
what is the distribution of degenerate irreducible representations that one
generates?  What is the most likely representation? Is this
representation a classical representation?  It is well known that the
$N_c\rightarrow \infty$ limit of QCD is a classical theory even for
$k=1$~\cite{Yaffe}.  What happens for finite $N_c$ but $k \gg
1$~\footnote{The former corresponds to the large $N$ limit where the
invariance group of the theory grows with $N$. Several such large $N$
cases have been discussed in Ref.~\cite{Yaffe}. We are instead
interested in a situation where the underlying symmetry group is
unchanged but where quantum operators in higher dimensional
representations appear.  Examples of these are the large spin limits
of quantum spin models~\cite{QuantSpin}.}?  We will demonstrate in the
following that, for an $SU(N_c)$ gauge theory, the problem can be
formulated as a random walk problem in the space spanned by the
$N_c-1$ Casimir operators of the group.

Our interest in the questions posed here arise in the context of a
particular model of small $x$ physics in QCD, the
McLerran-Venugopalan (MV) model for small $x$ parton
distributions in large nuclei~\cite{MV}. The MV model is a simple
model to understand the physics of saturation~\cite{GLR} at small $x$
in QCD. To highlight the physical relevance of the formal mathematics,
we will therefore address solutions to these general questions in the
specific context of the MV model.  However, the solution to the
mathematical problem posed may be of more general interest and will
hopefully prove useful in developing novel treatments for a variety of
physical situations in QCD. These include problems in jet
physics~\cite{Dokshitser}, finite temperature transport
theory~\cite{BlaizotIancu} and even percolation and string based
models of multi-particle production~\cite{BiroKnollNielsen,FDP}.

This paper is organized as follows. In the following section, we will
briefly review the McLerran-Venugopalan model.  We will focus in
particular on those aspects of the model related to the random walk of
partons in the space spanned by the Casimirs of their color group and
of their subsequent treatment as classical color charge distributions.
In section 3, we consider the case of the MV-model in an $SU(2)$ gauge
theory.  The problem here is especially simple since it can be mapped
into a one dimensional random walk problem of $k$ spins of
spin-$1/2$. The peak of the distribution is at a representation whose
dimension is of order O($\sqrt{k}$).  The distribution of
representations about this peak value is well approximated as an
exponential in the quadratic Casimir operator weighted by $k$, as
indeed assumed in the MV-model.

The ``real world" case of an $SU(3)$ gauge theory is discussed in
section 4. In this case, one can think of the random walk as occurring
in the space spanned by the third component of the isotopic spin $I_3$
and the hypercharge $Y$.  Determining the distribution of higher
dimensional representations generated by $k$ quarks when the random
walk is in more than one dimension ($SU(N_c)$ for $N_c\geq 3$) is
non-trivial and is the primary subject of this paper.  We will show
explicitly, that for a random walk of $k$ quarks in the fundamental
$SU(3)$ representation, the distribution of degenerate representations
is, as in the $SU(2)$ case, also peaked at a representation of order
O($\sqrt{k}$). The distribution of representations about the most
likely representation, however, cannot simply be expressed in terms of
an exponential of the quadratic Casimir operator alone; the argument
of the exponential has an additional term proportional to the cubic
Casimir operator weighted by $k^2$. The cubic term thus acts as a
perturbation of the quadratic term. The relative magnitude of the
former to the latter is of order O($1/\sqrt{k}$) with a coefficient
that is computed exactly. Thus for very large nuclei $A\rightarrow
\infty$, the contribution from this term can be
neglected~\footnote{The statements here are valid for values of
$x$ where small $x$ quantum evolution is not too
important. Quantum evolution effects will be discussed briefly later
in this section and in the final section.}. We also consider the
representations generated by random walks of gluons and of
quark-anti-quark pairs. Interestingly, in both of these cases, the
distribution of representations is given by an exponential in the
quadratic Casimir alone. This is true because the quadratic Casimir, unlike
the cubic Casimir, is symmetric in the upper and lower $SU(3)$ tensor indices
(denoted
by $m$ and $n$ in the paper).
Gluons and quark-anti-quark representations, unlike the quark representations,
favor representations that are symmetric in $m$ and $n$.

The generalization of these results to $SU(N_c)$ for any general $N_c$ is
straightforward and is briefly discussed in section 5.
The essential features of the $SU(3)$ result are preserved: the
distribution of representations about the most likely representation
is determined primarily by the quadratic Casimir. The other $N_c-2$
Casimirs add small (and in principle quantifiable) perturbative
corrections--all of which vanish for $k\rightarrow \infty$.

In section 6, we will summarize our results and discuss their
ramifications. A brief discussion of classical limits of quantum
systems in general, and quantum spin systems in particular, is
presented in appendix A. The other appendices contain technical
details of results presented in the body of the paper.

\section{Classical color charges in the  McLerran-Venugopalan model}
\vskip 0.1in

The McLerran-Venugopalan model is a classical effective field theory
for wee parton distributions in large nuclei~\cite{MV}. The model is
formulated in the light cone gauge $A^+=0$, and in the infinite
momentum frame, where the momentum of the nucleus, $P^+ \rightarrow
\infty$.  In this case, the physics of time dilation ensures that the
time scales for partons carrying a higher fraction $x$ of the nuclear
momentum (the ``valence" partons with $x\sim 1$) to interact with one
another is much larger than the characteristic time scale for their
interactions with softer ``wee" partons (with $x\ll 1$). The softer
partons, in turn, couple to a large number of valence partonic sources,
which appear static on the wee parton time scales. In particular, wee
partons with coherence lengths ($l_{\rm coh.}\approx 1/2 m_N x$) much
greater than the Lorentz contracted nuclear widths (of order
$2R/\gamma$), or equivalently with $x\ll A^{-1/3}$, couple coherently to
$\sim A^{1/3}$ valence partons along the longitudinal length of the
nucleus in the infinite momentum frame. Here $m_N$ is the nucleon mass
and $\gamma \sim P^+/m_N$ is the Lorentz factor in the infinite
momentum frame.

The separation of scales in $x$ between wee and valence partons is
experimentally well established in parton distribution function
measurements~\cite{ZEUSH1}. The insensitivity of the valence
distributions to the sea is also experimentally well established-the
well known phenomenon of ``limiting fragmentation'' is a direct
consequence~\cite{Phobos}.

Further, on the time scales relevant to wee parton dynamics, it is
reasonable to assume that the valence parton sources are random light
cone sources. This is plausible because firstly, most of the hard
partons are confined in different nucleons (and hence do not interact
with one another); secondly, due to time dilation, even hard partons in the
same nucleon are
non-interacting (and therefore independent sources) over the short
time scales relevant to the wee partons.

How many of these random sources the wee partons actually couple to
depends on the typical transverse momentum of the wee
parton~\footnote{The wee parton is soft only in its longitudinal
momentum-its transverse momentum may be large.}. A wee parton with
momentum $p_\perp$ resolves an area in the transverse plane $
(\Delta x_\perp)^2 \sim 1/p_\perp^2$. The number of valence partons it
interacts
simultaneously with is then
\be
k \equiv k_{(\Delta x_\perp)^2} = {N_{\rm valence}\over \pi R^2}\, (\Delta
x_\perp)^2
\, ,
\label{eq:quarkno}
\ee
which indeed is proportional to $A^{1/3}$ since $N_{\rm
valence}=3\cdot A$ in QCD.  This counting of color charges is only
valid as long as $p_\perp > \Lambda_{\rm QCD}$ since, on the scale of
the nucleon size , $p_\perp \sim \Lambda_{\rm QCD}\sim 200$ GeV,
confinement ensures that the wee partons see no net color charge.  The
momentum scale for color neutrality may be even larger than the
confinement scale due to screening effects analogous to Debye
screening~\cite{IIM,Mueller}. These effects, while very interesting,
are beyond the scope of this paper.

It was argued in the MV model that when $k\sim A^{1/3} \gg 1$, the
most likely color representation that the wee partons couple to is a
higher dimensional representation. We will demonstrate later that this
representation is one of order $O(\sqrt{k})$. Thus, as discussed
further in appendix A, for large enough $k$, the color charge
distribution can be treated as a classical color charge
distribution. It is further argued in the MV-model, that the
distribution of classical representations is Gaussian, with a variance
proportional to $k$ (or, equivalently, to $A^{1/3}$).

With these stated assumptions, the classical effective Lagrangian for the
MV-model,
formulated in the infinite momentum frame ($P^+\rightarrow \infty$) and light
cone gauge ($A^+=0$) has the form
\be
{\cal L} = \int d^4 x \left[ {1\over 4}F^a\,F^a - J\cdot A \right]
+ i\int d^2 x_t {\rho^a \rho^a \over 2\mu_A^2} \,.\nonumber \\
\label{eq:one}
\ee
The first term is the usual QCD Field Strength tensor squared (Lorentz indices
are suppressed for simplicity here), which describes the dynamics of the wee
partons.
The second term denotes the
coupling of these wee partons to the hard valence parton sources. The valence
parton
current has the form
\be
J^{\mu,a} = \rho^a(x_t)\delta(x^-)\delta^{\mu +} \, ,
\label{eq:current}
\ee
where $\rho$ is the {\it classical} color charge per unit transverse
area of the valence parton sources~\footnote{As written, this term in
the action is not gauge invariant. Gauge invariant expressions are
given in Refs.~\cite{JKLW,JSR}. The lowest order in $g$ term in the
expansion of these expressions gives the form shown here.}.  The
likelihood of configurations of differing $\rho$'s is given by the
final term in Eq.~(\ref{eq:one}) -- the weight $\mu_A^2$ of the
Gaussian is the average color charge squared per unit area per color
degree of freedom.  The correlator of color charge densities is
then~\footnote{There are several conventions for the color charge
densities in the literature. For a discussion, see Ref.~\cite{KNV2}.}
\be
\ave{\rho^a(x_\perp)\rho^b(y_\perp)}
= \mu_A^2\, \delta^{ab}\,\delta^{(2)}(x_\perp -y_\perp) \, .
\label{eq:correlator}
\ee

The average color charge squared per unit area per color degree of
freedom, $\mu_A^2$ is simply determined in the MV-model. The color
charge squared in a tube of transverse area $(\Delta x_\perp)^2$ is the
color charge squared per unit quark ($g^2 C_F$) times the number of
quarks in the tube: $(\Delta x_\perp)^2 \cdot A\,N_c/\pi R^2$. When this is
normalized as in Eq.~(\ref{eq:correlator}), per unit transverse area,
per color degree of freedom, one obtains
\be
\mu_A^2 = {g^2 A\over 2\pi R^2} \,\,,
\label{eq:mua2}
\ee
which is of order $A^{1/3}$. For a very large nucleus, $\mu_A^2 \gg
\Lambda_{\rm{QCD}}^2$, and since it is the only scale in the problem,
the coupling constant $\alpha_s\equiv \alpha_s(\mu_A^2)\ll 1$. Since
the coupling is weak, one can compute parton distributions for a large
nucleus. The classical field equations can be solved and it was shown
explicitly that the number distribution is of order ~$1/\alpha_S$ for
$p_\perp^2 < g^2 \mu_A^2$~\cite{JKMW}.

The Gaussian functional weights and the classical field equations for
a large nucleus are also recovered in a particular model of large
nuclei where the nucleons are modeled as color singlet quark
anti-quark pairs~\cite{Kovchegov96}.  In Refs.~\cite{JKMW} and
{}~\cite{Kovchegov96}, it was recognized that it was essential to
smear the $\delta$-function sources in Eq.~(\ref{eq:current}) in the
$x^-$ direction to obtain regular classical solutions.  Quantum
corrections to the MV-model are large~\cite{AJMV} but they can be
absorbed, via a Wilsonian renormalization group procedure, at each
step in $x$ into new sources and fields, while preserving the
essential structure of the classical effective
Lagrangian~\cite{JKLW,CGC}.  Thus this simple MV-model can be
reformulated as a more sophisticated effective field theory, the
``Color Glass Condensate'' (CGC), which is applicable to both hadrons
and nuclei. In the CGC, the weight functional can be represented more
generally as
\be
\exp\left(-W[\rho]\right) \,,
\label{eq:two}
\ee
where $W[\rho]$ is a gauge invariant functional that satisfies the
non-linear renormalization group (RG) equation in $x$. For a recent
review of the MV model and the CGC, see Ref.~\cite{IV}.

In the rest of the paper, we will restrict ourselves to the MV model and seek
to establish more rigorously the
following, namely,
\begin{itemize}
\item  the most likely representation the wee partons couple to is a higher
dimensional representation which
can be represented in terms of a classical color charge density $\rho$,
\item  the sum over color charges in the path integral can be expressed as a
path integral over classical
color charges, with a weight that is well approximated by a Gaussian for a
large number $k$ of color charges. This is shown for quark sources as well as
for
gluon and quark--anti-quark sources for $N_c\geq 3$.
\item Finally, we will discuss, for  $N_c\geq 3$, possible sub-leading
corrections to the Gaussian term and
their relative dependence on $k$.
\end{itemize}

\section{Random walks and classical color charge representations in two color
QCD at small $x$}
\vskip 0.1in

We now consider the case of the MV-model in $SU(2)$ QCD. As discussed
previously, (see Eq.~(\ref{eq:quarkno})) the wee partons couple to a
large number of uncorrelated quarks-in this case, quarks in the
fundamental $SU(2)$ representation. We wish to discuss here the
distribution of representations generated by adding $k$-quarks.

\subsection{Random walk of many spin-1/2 quarks}

The problem of adding $k$ random color charges in $N_c=2$ QCD is exactly
equivalent to the problem of adding $k$ spins of
spin-1/2--they both correspond to an internal symmetry group whose generators
are elements of the $SU(2)$ algebra.
Therefore, let's start with a spin 0 singlet state.  Multiplying with a spin
$1/2$
state results in a $1/2$ state, or
\be
 \bbox{2 \times 1 = 2}
\ee
where $\mbf{1}$ is the singlet state and $\mbf{2}$ denotes the spin 1/2
state.  In general we denote a spin $l$ state
by the degrees of freedom associated with the state, namely,
$\bbox{s} = 2l+1$.

Continuing, multiplying another spin $1/2$ states results in
 \be
 \bbox{2 \times 2 = 1 + 3}
 \ee
 If we have 3 spin 1/2's, we get
 \be
 \bbox{
 2 \times 2 \times 2
 =
 2\times (1 + 3)
 =
 2 + 2 + 4 \, ,
 }
 \ee
and so on.

Let $v_s^{(k)}$ denote the multiplicity of the representation $\mbfs$ when $k$
fundamental representations are multiplied.
 In general, if one adds one more spin 1/2 particle to an $\bbox{s}$
 state, the result is a mixture of the states $\bbox{s-1}$ and
 $\bbox{s+1}$:
 \be
 \bbox{2 \times s = (s-1) + (s+1)}
 \label{eq:2timesn}
 \ee
 Therefore, the multiplicity of $\bbox{s}$ state in the $k$-th iteration
 is given by
 \be
 v_s^{(k)} = v_{s-1}^{(k-1)} + v_{s+1}^{(k-1)}
 \label{eq:spin_half}
 \ee

We wish to find the solution of this recursion relation with the initial
condition
\be
 v^{(0)}_0 = 1\ \  \hbox{otherwise}\ \   v^{(0)}_s = 0 
\label{eq:init_conds}
\ee
We begin by noting that
the binomial coefficients satisfy a similar (but not identical) recursion
relation,
 \be
 \left( \colvec{k\\s} \right)
 =
 \left( \colvec{k-1\\s-1} \right)
 +
 \left( \colvec{k-1\\s} \right) \, .
\label{eq:binomial}
\ee
Therefore,  we can try and find a solution to Eq. (\ref{eq:spin_half})
in terms of binomial coefficients.

Define
 \be
 G_{k:\,s} =
 \left(\colvec{k\\(k+s)/2}\right) \, .
\label{eq:Gbinom}
 \ee
which is non-zero only if $(k+s)$ is even.
Consider now the combination
\be
g(k:s) & = & G_{k:\,s-1} - G_{k:\,s+1}
\non
& = &
 \left(\colvec{k\\(k+s-1)/2}\right) \,
 -
 \left(\colvec{k\\(k+s+1)/2}\right) \, .
\ee
Using the binomial coefficient relation (\ref{eq:binomial}), one can easily
show that this combination satisfies the recursion relation
\be
g(k:s) = g(k-1:s-1) + g(k-1:s+1)
\ee
Furthermore, for $k=0$
\be
g(0:1)
 =
 \left(\colvec{0\\0}\right)
 -
 \left(\colvec{0\\1}\right)
 = 1
\ee
and all the other $g(0:s)$'s are zero.

In this way, we have found
a solution to the recursion relation (\ref{eq:spin_half})
satisfying the right boundary condition, Eq.~(\ref{eq:init_conds}), namely,
\be
v^{(k)}_{s}= g(k:s) = G_{k:\,s-1} - G_{k:\,s+1} \,
\label{eq:bin-soln2}
\ee

Since we
know the $G$'s explicitly in terms of binomial coefficients
(Eq.~(\ref{eq:Gbinom})), we can
determine $v^{(k)}_{s}$ explicitly.  Since we are interested in the limit $k\gg
s\gg 1$,  we can use the well known Stirling formula to approximate
\be
G_{k:\,s} &=& {k!\over ( (k+s)/2 )!\, ( (k-s)/2 )!} \, ,
\ee
as
\be
\ln G_{k:\,s}\approx  k \ln 2 - s^2/2k - (1/2) \ln k - (1/2) \ln(2\pi)\,\,,
 \ee
or equivalently,
\be
G_{k:\,s} \approx {2^{k-1/2}\over \sqrt{k\pi}} e^{-s^2/2k} \, .
 \ee

Thus the solution of the recursion relation for $k \gg s \gg 1$ is
\be
v^{(k)}_{s} & = &
 G_{k:\,s-1} - G_{k:\,s+1}
 \non
 & \approx &
 -2 \partial_s G_{k:\, s}
 \non
 & \approx & {2^{k+1/2}\over k\sqrt{k\pi}}\,s\, e^{-s^2/2k}
 \label{eq:vks}
 \ee
This formula represents
the distribution of representations with spin $l = (s-1)/2$
when there are $k$ random
spin 1/2 particles in the system.

To find the {\em probability} of the spin state $l$,
we need to take into account the degeneracy of the spin state $s = 2l+1$.
Since we have $k$ random spin 1/2 particles, summing the degeneracy times
the multiplicity ($s\, v^{(k)}_s$) over
all possible representations must be equal to the  $2^k$ degrees of freedom.
Even with our large $k$ approximation, we find that this  is indeed the case:
\be
\int_0^\infty ds\, s \, v^{(k)}_s
 = 2^{k} \, .
\label{eq:2k}
\ee

Defining the ``classical'' spin vector~\footnote{We will see {\it a posteriori}
that $\bfl^2$ is of order $k$. For
large $k$, we argue in Appendix A that such representations are classical
representations.} $\bfl$ which is related to $s$ by the relation
\be
s^2 = 4\,\bfl^2 \, ,
\ee
or equivalently $\bfl^2 = (l+1/2)^2$,
we can exploit the lack of angular dependence in the argument to re-write
Eq.~(\ref{eq:2k}) as
\be
1= \left(2\over k\pi\right)^{3/2}\, \int d^3 l \exp\left( - 2\,\bfl^2/k\right)
\,
{}.
\label{eq:su2norm}
\ee
Thus one can define
\be
{\cal P}_k({\bfl}) =
\left(2\over k\pi \right)^{3/2}\, e^{-2{\bfl^2}/k}\,\,,
\label{eq:su2prob}
\ee
to be the probability density for the system of $k$ spin-1/2
particles to have the total spin $\bfl$.
The above formula clearly resembles the classical Maxwell-Boltzmann
distribution of particles in a heat bath.

Our result can be easily expressed in terms of the $SU(2)$
Casimir, which is
\be
D_2 = l(l+1) \approx {\bfl}^2 \, .
\ee
The most probable value of the Casimir is then given by
\be
\bar{D}_2 = {k\over 4}
\ee
or $l = \sqrt{k}/2 + O(1)$.

One can repeat the same analysis for the case of many spin-1
particles. In this case, one has a somewhat more complex analysis
(involving trinomial coefficients) but the final result for the
multiplicity distribution is identical to the spin-1/2 case-up to
overall constants.

What does this result for $v_s^{(k)}$ imply for the derivation of the
effective action in the MV model? In order to understand this, we have
to step back a little and discuss, with some greater detail, the
derivation of the MV-effective action~\cite{MV}.

\subsection{From random walks to path integrals}

The path integral describing the ground state $\ket{O}$
of a large nucleus can be written as
\be
{\cal Z} = \bra{O}e^{i\,x^+\,P_{\rm QCD}^-}\ket{O}
= \lim_{x^+\rightarrow i\infty}
\sum_{N,Q} \bra{N, Q}e^{i\,x^+\,P_{\rm QCD}^-}\ket{N, Q} \, ,
\label{eq:pathintegral_su2}
\ee
The sums over $N$ and $Q$ here respectively represent the sum over all
possible states in the path integral and a sum over the color quantum
numbers of these states~\footnote{The distinction between the quantum
numbers represented by $N$ and the color degrees of freedom
represented by $Q$ is of course artificial. We make this distinction
because it will prove useful in constructing the effective
theory.}. In the lattice representation of the path integral, a state
$\ket{N, Q}$ would correspond to a 4-dimensional box on the lattice
containing a net color charge $Q$. In the MV-model, one is interested
in a coarse grained theory, where the size of the box is chosen (see
the first paper in Ref.~\cite{MV}) such that it contains a large
number of quark color charges, namely, $k \gg 1$.
{}From
Eq.~(\ref{eq:quarkno}), this implies that the coarse grained state
$\ket{N,Q}$ contains only modes with $\Lambda_{\rm QCD}\ll p_\perp^2
\ll \mu_A^2$ (see Eq.~(\ref{eq:mua2})) and the corresponding color
charge $Q$ is the charge constructed from all the $k_{(\Delta x_\perp)^2}$
quarks contained in a box with transverse size
$1/\mu_A^2 \ll (\Delta x_\perp)^2 \ll 1/\Lambda_{\rm QCD}^2$.

In general, one has to perform the quantum mechanical sum over the
color charges to construct $\ket{N,Q}$, which is a difficult problem
indeed. It is at this point one sees the relevance of the problem of
the distribution of representations that we have just solved (for the
$N_c=2$ case). In the MV model, the color charges are random, and $k$
charges can be distributed in a wide range of distributions. Since the
most likely representation is ${\bar s}= \sqrt{k}/2 \gg 1$ for $k\gg
1$, the color charge corresponding to this representation, as argued
in appendix A, is a classical representation.

Thus in the $SU(2)$ case, the sum over all spin states in the path
integral can be replaced by the integral~\footnote{This assumes that
the $s=2l+1$ states in a representation have the same $P^+$.}
\be
\sum_{l} v_l^{(k)} \sum_{m=-l}^l \ket{l, m}\bra{l,m}
\to
\int d^3 l\, e^{-2{\bfl}^2/k}\, ,
\label{eq:su2prob1}
\ee
where we have made use of Eqs.~(\ref{eq:vks}) and (\ref{eq:su2prob}).
For a transverse area of size $(\Delta  x_\perp)^2$ containing $k$ spins,
then one can introduce classical color charge (spin)
density $\rho^a$ which is defined by the relation,
\be
l_a = {1\over g} \int_{(\Delta x_\perp)^2} d^2 x_\perp \, \rho_a(x_\perp)
\approx
{(\Delta x_\perp)^2\over g} \rho_a(x_\perp) \, .
\label{eq:su2charge1}
\ee
We can now re-express the Gaussian weight in Eq.~(\ref{eq:su2prob1})
in terms of the classical color charge density:
\be
2\,{{\bfl}^2\over k}
& = &
N_c\, {{\bfl}^2\over k}
\non
& =&
N_c \,{(\Delta x_\perp)^4 \over g^2 k} \rho_a \rho_a
\non
& = &
{\pi R_A^2\over g^2 A}\, (\Delta x_\perp)^2\, \rho_a \rho_a \, .
\label{eq:su2charge2}
\ee
Here we have used Eq.~(\ref{eq:quarkno}) to express $k$ in terms of
the number of valence quarks in an area $(\Delta x_\perp)^2$.  We also
used the fact that the number of valence quarks per baryon is equal to
$N_c$.

The sum in Eq.~(\ref{eq:su2prob1}) can therefore be expressed as
\be
\int \prod_{a} d\rho^a\, W[\mbf{\rho}]
\equiv
\int \prod_a d\rho^a\,
\exp\left(-\int_{A} d^2 x_\perp\,
{\mbf{\rho}(x_\perp){\cdot}\mbf{\rho}(x_\perp)\over 2\mu_A^2}
\right)
\label{eq:MVsu2}
\ee
where
\be
\mu_A^2 = {g^2 A \over 2 \pi R_A^2}
\label{eq:muasu2}
\ee
is independent of $N_c$, and is the color charge squared per unit area defined
in the MV-model (Eq.~(\ref{eq:mua2})).
Integrating over all the boxes of area $(\Delta x_\perp)^2$ in the transverse
plane, one obtains the path integral for classical color charge distributions
in the MV-model~\cite{MV}.

\section{Random walks and classical color charge representations for $N_c=3$
QCD at small $x$}

We shall now extend our discussion in the previous section to the case
of three color QCD. The problem is more non-trivial here since one now
has a 2-dimensional random walk in the space generated by the two
$SU(3)$ Casimirs.  Thus it is no longer obvious that the quadratic
Casimir alone will contribute to the ``Boltzmann" weight in the path
integral over color charges.  We will show in this section that the
approximation that only the quadratic Casimir contributes is an
excellent approximation when the number of quarks is large.  We will
first consider the case where the sources of color charge are $SU(3)$
quarks only before considering the representations generated by random
gluons and quark-anti-quark states respectively.

\subsection{Many $SU(3)$ quarks}

\begin{figure}[thb]
\epsfxsize=8cm
\epsfbox{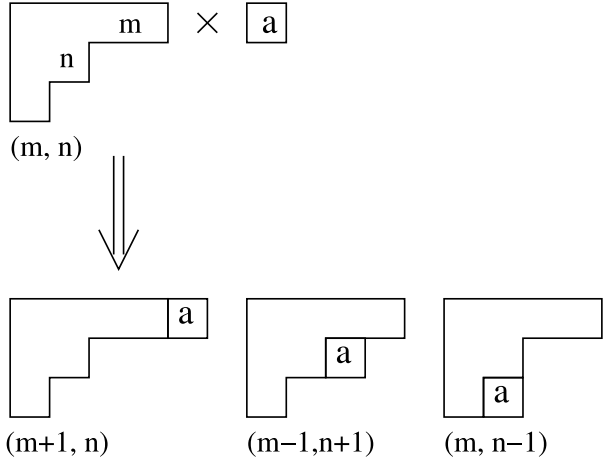}
\caption{Young tableaux representing the result of multiplying
$SU(3)$
quarks in a
representation labeled by the tensor indices m \& n with a quark in the ${\bf
3}$ representation. }
\label{fig:Young1}
\end{figure}

In the $SU(3)$ case,
representations are labeled by  two integers $m$ and $n$
and written as $(m,n)$.
Here $m$ and $n$ are the number of the upper and the tensor indices,
respectively.
A  fundamental quark ${\bf 3}$ state
in $SU(3)$ is hence $(1,0)$ while an anti-quark in
the ${\bf {\bar 3}}$ state is $(0,1)$.
 As is well known, $SU(3)$ representations can be conveniently represented
graphically as Young tableaux, where
 $m$ denotes the number of boxes in
 the uppermost row minus those in the middle row,
while  $n$ is the number of boxes in the middle row minus that of the bottom
row.

Recall that
we are interested in studying the distribution of representations when
one adds $k$ quarks in the fundamental
representation. Consider what happens when one adds a quark to an $(m,n)$
representation.  This is
represented by the Young tableaux in Fig.~\ref{fig:Young1} which allows us to
deduce that
 \be
 \bbox{
 (1, 0) \times (m, n)
 =
 (m+1, n) + (m-1, n+1) + (m, n-1)
 }
 \label{eq:quarks}
 \ee

 We shall represent the multiplicity of each state by a matrix element
 $N_{m, n}$
 with the indices $0 \le m, n$ representing the same $(m, n)$ of the
 representations. (Note that these $N_{m,n}$ are the $SU(3)$
 analogs of the $v_s$
in the $SU(2)$ case.)
 The multiplicity of the $(m, n)$ state in the ($k+1$)-th iteration is then
 given by the recursion relation
 \be
 N^{(k+1)}_{m,n} =
 N^{(k)}_{m-1,n}
 +
 N^{(k)}_{m+1,n-1}
 +
 N^{(k)}_{m,n+1}
 \label{eq:su3_q_rec}
 \ee
 for $m, n \ge 0$ with the understanding that
 $N^{(k)}_{m,-1} = N^{(k)}_{-1,n}=0$.
 The initial condition is
 \be
 &
 N^{(0)}_{0,0} = 1 \ \ \hbox{otherwise}\ \ N^{(0)}_{m,n} =0
 \label{eq:su3_q_init}
 \ee
As in the $SU(2)$ case, we now wish to determine $N^{(k)}_{m,n}$ by solving
this
recursion relation.

 We begin by noting that Eq.~(\ref{eq:su3_q_rec}) represents a trifucation
process since we are multiplying $\bbox{3}$'s.  The basic unit
should therefore be
the tri-nomial coefficients~\footnote{Trinomial coefficients also arise in the
adjoint spin-1 representation of $SU(2)$.}:
 \be
 C_{k:\; l_1, l_2, l_3}
 \equiv
 {k!\over l_1!\, l_2!\, l_3!}\delta_{k-l_1-l_2-l_3}
 \ee

 Following the pattern set by \equ{eq:Gbinom} and \equ{eq:quarks}, we define
 \be
 (m, n)
 =
 l_1 (1, 0) + l_2 (-1, 1) + l_3 (0, -1) \, ,
 \ee
Re-defining $C_{k\,;l_1,l_2,l_3}$ in terms of $m,n$ as $G_{k;m,n}$, we have,
 \be
G_{k:\; m, n} =
 {k!\over
 \left(k+2m+n\over 3\right)!
 \left(k-m+n\over 3\right)!
 \left(k-m-2n\over 3\right)!} \, .
 \ee
 We are now in a position to compute the multiplicity of representations.
 One finds that
 \be
 N_{m,n}^{(k)}
 & = &
 G_{k:\; m, n} + G_{k:\; m+3, n} + G_{k:\; m, n+3}
 \non
 & & {}
 -
 G_{k:\; m+2, n-1} - G_{k:\; m-1, n+2} - G_{k:\; m+2, n+2}
 \label{eq:su3_q_sol}
 \ee
 satisfies the recursion relation Eq.~(\ref{eq:su3_q_rec}) and the initial
 condition Eq.~(\ref{eq:su3_q_init}).  We have verified this analytically and
 have checked numerically as well that the recursion relations
 (\ref{eq:su3_q_rec}) and our solution (\ref{eq:su3_q_sol}) indeed yield
 the same values upon iteration.

As performed previously in the $SU(2)$ case,
we can use the Stirling formula to
write the above expression for $G$
(after some algebra) compactly as
 \be
 \ln G_{k:\; m, n}
 \approx
 \ln\left(3^{k+3/2}\over 2 \pi k \right)
 - {(m^2 + mn + n^2)\over k} + {(m - n) (2 m + n) (m + 2 n)\over 6k^2}
 \ee
 for $k \gg m, n \gg 1$.

 Recall that the $SU(3)$
 Casimir for a given $m, n$ state is given by~\cite{Close}
 \be
 D_2^{(m,n)} =
 {1\over 3}\left(m^2 + mn + n^2\right) + (m + n) \, .
 \ee
Note further that this Casimir is symmetric in $m$ and $n$. This is unlike the
Cubic
Casimir which is anti-symmetric under exchange of $m$ and $n$ and can be
written as~\cite{Macfarlane}
\be
D_3^{(m,n)} = {1\over 18}
\left( m + 2 n + 3 \right)
\left( n + 2 m + 3 \right)\left(m - n\right)
\ee
A little algebra should suffice to convince ourselves that $G$ can be
simply written in terms of the $SU(3)$ Casimirs as~\footnote{Here we
have kept only the leading terms in $D_2$ and $D_3$ and expanded the
exponential in $D_3$ to lowest order. Our approximation can be checked
to be self-consistent.}
 \be
 G_{k:\, m, n}
 & \approx &
 {\frac{{3^{{\frac{3}{2}} + k}}}{2\,k\,\pi }}
 \, \exp\left(-3 D_2^{m,n}/k \right)
 \left( 1 + 3 D_3^{m,n}/k^2\right)\, .
 \label{eq:cubic}
 \ee
Clearly, near the peak, one has $m = O(\sqrt{k})$ and $n =
O(\sqrt{k})$.  Hence the cubic Casimir introduces a correction of size
$O(1/\sqrt{k})$ for large k.

In the large $k$ limit, one finds that the multiplicity can be approximated
as
 \be
 N_{m,n}^{(k)}
 & = &
 G_{k:\; m, n} + G_{k:\; m+3, n} + G_{k:\; m, n+3}
 \non
 & & {}
 -
 G_{k:\; m+2, n-1} - G_{k:\; m-1, n+2} - G_{k:\; m+2, n+2}
 \non
 & \approx &
 2\, \partial_m^3 G_{k:\, m, n}
 +
 2\, \partial_n^3 G_{k:\, m, n}
 -
 3\, \partial_m \partial_n^2 G_{k:\, m, n}
 -
 3\, \partial_n \partial_m^2 G_{k:\, m, n}
 \non
 & \approx &
 {27\,m\,n\,\left( m + n \right) \over k^3}\,
 {\frac{{3^{{\frac{3}{2}} + k}}}{2\,k\,\pi }}
\, \exp\left(-3\, D_2^{m,n}/k \right)\,,
 \label{eq:approx_33}
 \ee
where we have dropped the $D_3$ contribution.

Recall that the dimension of an SU(3) representation is
 \be
 d_{mn} = {(m+1)(n+1)(m+n+2)\over 2} \approx {mn(m+n)\over 2}\, .
 \ee
Therefore, just as in the SU(2) case, the probability to find the
system of $k$ quarks in an $(m,n)$ representation should be
proportional to $d_{mn}\, N^{(k)}_{m,n}$.  Furthermore, one should
recover the total number of degrees of freedom $3^k$ by summing
$d_{mn}\, N^{(k)}_{m,n}$ over all possible representations.

With the expression obtained so far,
one can show that  integrating $N_{m,n}^{(k)}$ weighted by $d_{mn}$ over
 $0 \le m, n \le \infty$ yields,
 \be
 {\frac{27 \times {3^{{\frac{3}{2}} + k}}}{4\,k^4\,\pi }}
 \int_0^\infty dm dn\,
 {m^2\,n^2\,\left( m + n \right)^2}\,
\, \exp\left(-3 D_2^{m,n}/k \right)
 = 3^{k+1}
 \ee

The discrepancy with the integral above arises from the fact that the
 actual $N^{(k)}_{m,n}$ is rather asymmetric in $m$ and
 $n$~\footnote{Our method yields the correct number of degrees of
 freedom for gluons and quark-anti-quark pairs which, unlike quarks,
 are symmetric in $m$ and $n$ (see sections 4.2 and 4.3).}.  Since we
 are multiplying $k$ number of $\bbox{3} = (1,0)$ representations, the
 $m$ side is more heavily populated than the $n$ side.  On the other
 hand, the approximation given in Eq.~(\ref{eq:approx_33}) is symmetric
 in $m$ and $n$ which is valid only in the vicinity of the peak.

Nevertheless, our result enables us to normalize the distribution of
representations as
 \be
 1 =
 {\frac{27 {\sqrt{3}}}{4\,k^4\,\pi }}
 \int_0^\infty dm dn\,
 {m^2\,n^2\,\left( m + n \right)^2}\,
\, \exp\left(-3 D_2^{m,n}/k \right)
 \label{eq:total_dof1}
 \ee

 We shall show below  that we can re-write Eq.~(\ref{eq:total_dof1}),
 in analogy to Eq.~(\ref{eq:su2norm}), as
 \be
 1 = \left(N_c\over k\pi \right)^{4}\int d^8 Q\, e^{-N_c\,
\bfQ^2/k}
 \label{eq:total_dof2}
 \ee
 where $\bfQ = (Q_1, Q_2, \cdots, Q_8)$
 is a classical color charge vector defined by $|\bfQ|=\sqrt{Q^aQ^a} \equiv
\sqrt{D_2^{m,n}}$
 and $Q_1,\cdots,Q_8$ are its eight components. Once we are able to do this,
the
rest of the argument, expressing
 the sum over color charges in large nuclei as a path integral over Gaussian
color charges, will follow in smooth
 analogy to the derivation in section 3.2.

For any $SU(3)$
representation (not necessarily a classical representation), one
can write
the measure $d^8 Q$ as~\cite{Shatashvilli,Johnson,Kelly}
\be
 d^8 Q
 =
 d\phi_1
 d\phi_2
 d\phi_3
 d\pi_1
 d\pi_2
 d\pi_3
 dm \, dn
 \,
 \left( m n (m + n) {\sqrt{3}\over 48} \right)\, ,
\label{eq:darboux1}
\ee
where $\phi_i$ and $\pi_i$  (the so called ``Darboux" variables) are
canonically
conjugate variables~\footnote{It can be
checked that $Q_1\cdots Q_8$ satisfy $\left\{Q_a,Q_b\right\}_{\rm PB} = f_{abc}
Q_c$ where the Poisson Bracket is defined
as $\left\{Q_a,Q_b\right\}_{\rm PB} = \sum_i^3 \left({\partial Q_a\over
\partial \phi_i}{\partial Q_b\over\partial \pi_i}-{\partial Q_a\over \partial
\pi_i}{\partial Q_b \over \partial\phi_i}\right)$.}.
If the integrand depends only on $m$ and $n$, we can carry out the integral
over the angles $d\phi_i$ and their
canonically conjugate momenta $d\pi_i$. One obtains (see appendix B for
details of the derivation)~\cite{Johnson}
\be
\int \prod_{i=1}^3 d\phi_i \,d\pi_i =
{(2\pi)^3\over 2}\,m n (m+n) \,
{}.
\label{eqn:darboux2}
\ee

With this result, we can write Eq.~(\ref{eq:darboux1}) as
\be
\int d^8 Q  = {(2\pi)^3 \over 32\sqrt{3}}
\int dm\, dn \left( m^2  n^2 (m + n)^2 \right) \, .
\label{eq:darboux3}
\ee
Substituting Eq.~(\ref{eq:darboux3}) into
Eq.~(\ref{eq:total_dof1}), we obtain Eq.~(\ref{eq:total_dof2}).

In summary,
we have shown in this section  that for a sum over classical $SU(3)$
representations, the measure can be
expressed as an eight dimensional integral over the
components of the classical color
charge $\bfQ$ (with a magnitude of order $\sqrt{k} \gg 1$).

\subsubsection{From random walks to path integrals in $SU(3)$ QCD}

The reader may note that Eq.~(\ref{eq:total_dof2})
is exactly analogous to Eq.~(\ref{eq:su2norm})
obtained for $SU(2)$ quark sources. Then replacing $\bfl\rightarrow \bfQ$ in
Eqs.(\ref{eq:su2prob1}), (\ref{eq:su2charge1})
and (\ref{eq:su2charge2}), we recover
\be
\int \prod_{a}
d\rho^a\, W[\mbf{\rho}]
\equiv
\int \prod_a d\rho^a\,
\exp\left(-\int_{A} d^2 x_\perp\,
{\mbf{\rho}(x_\perp){\cdot}\mbf{\rho}(x_\perp)\over 2\mu_A^2}
\right)
\ee
where now $a=1,\cdots, 8$ and
\be
\mu_A^2 = {g^2 A\over 2 \pi R_A^2 }\, ,
\ee
is the color charge squared per unit area and is independent of $N_c$ as stated
in the
previous section.
Thus, even in the $SU(3)$ case, and as conjectured in the MV model, the leading contribution 
to the path
integral measure over classical color charge densities has a Gaussian weight
proportional to the quadratic Casimir.
As shown in Eq.~(\ref{eq:cubic}),
the contribution of the cubic Casimir to the weight is $O(1/\sqrt{k})$
which vanishes when $k\rightarrow \infty$.

\subsection{Many $SU(3)$ gluons}

In the case of uncorrelated gluons, we are multiplying the ${\bf 8}$'s.
Since the
Young tableaux for one $\bbox{8}$ has three boxes, we
add three
boxes at a time as we build up higher dimensional representations. As a
consequence, there
are sites in the $(m,n)$ plane we will never visit in the random walk. One can
see this as
follows. In general, a Young tableaux diagram for a state $(m,n)$ will have
$m+2n+3p$ boxes,
where $p$ is an integer with $p\geq 0$. Therefore, since we are multiplying
boxes in
multiples of 3, unless $m+2n=3l$ (where again, $l$ is an integer $\geq 0$), the
site $(m,n)$
can never be reached from the origin. In general, the allowed states in the
$(m,n)$ plane
have the form $(3s+u,3t+u)$, where $s,t$ are integers $\geq 0$, and $0\leq
u\leq 2$. Thus
for example, the state $(1,1)$ is allowed but not $(1,2)$ or $(1,0)$.

\begin{figure}[htb]
\begin{minipage}{8cm}
\psfig{figure=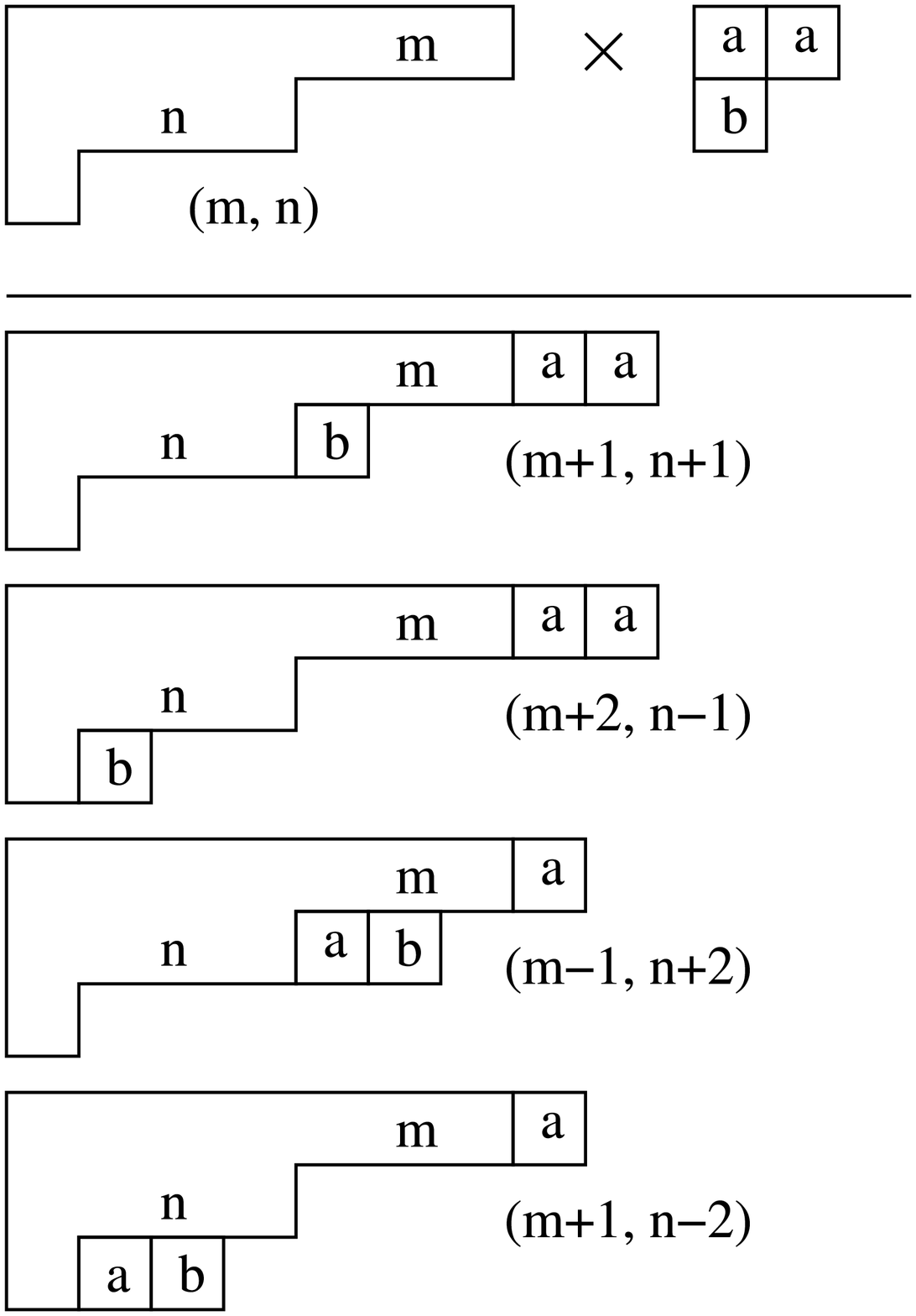,width=6cm}
\end{minipage}
\begin{minipage}{8cm}
 \psfig{figure=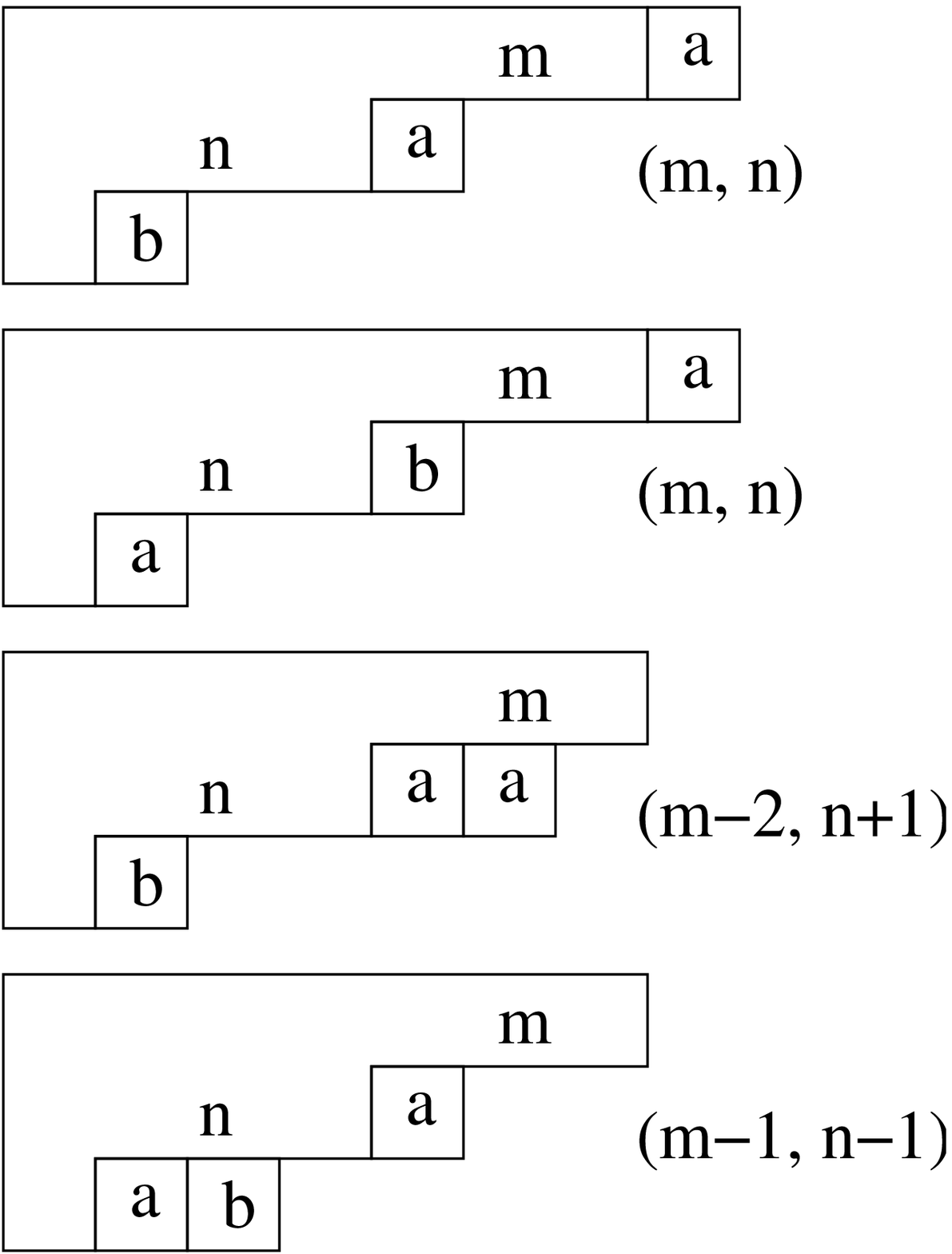,width=6cm}
\caption{Young tableaux denoting the result of the multiplication of an
arbitrary $SU(3)$ representation labeled by (m,n) with an ${\bf 8}$
representation }
\end{minipage}
\end{figure}

Fig.~2 shows Young tableaux  illustrating the representations generated by
multiplying an arbitrary $SU(3)$ representation $(m,n)$ with a $(1,1)$ (or
${\bf
8}$) representation. It is clear from the multiplication of Young tableaux that
one obtains
 \be
 \lefteqn{
 \bbox{
 (1, 1)\times (m, n)}} &&
 \non
 & = &
 \bbox{
 (m+1, n+1) +
 (m+2, n-1)
 }
 \non
 & &
 \bbox{
 {} +
 (m+1, n-2) +
 (m-1, n+2)}
 \non
 & &
 \bbox{
 {} +
 (m, n) +
 (m, n) +
 (m-2, n+1) +
 (m-1, n-1)
 }
 \non
 \label{eq:gluons}
 \ee
Just as in the case of many quarks we dealt with previously, we will denote the
multiplicity of a given state $(m, n)$ by a matrix
 $N_{m, n}$ with $m, n \ge 0$.
 Our initial state matrix is
 \be
 N^{(0)}_{0,0} = 1\ \ \ \hbox{otherwise}\ \ N^{(0)}_{m,n} = 0
 \ee
 From Eq.~(\ref{eq:gluons}),
 one can deduce immediately that the multiplicity of
 states is given by recursion relation,
 \be
 N^{(k+1)}_{m,n}
 & = &
 2\, N^{(k)}_{m,n}
 + N^{(k)}_{m+2, n-1}
 + N^{(k)}_{m+1, n+1}
 \non
 & & {}
 + N^{(k)}_{m+1, n-2}
 + N^{(k)}_{m-1, n+2}
 + N^{(k)}_{m-1, n-1}
 + N^{(k)}_{m-2, n+1}
 \label{eq:glue_rec1}
 \ee
Note that only states with $m$ and $n$ $\geq 0$ are allowed, so in the above,
the
multiplicity of states whose labels are negative is zero.

There are a few special cases which do not satisfy the above recursion
relation. These
are the states that lie on the edges of the first quadrant,
namely, $(m > 0, 0)$ and
$(0,n > 0)$ and at the origin $(0,0)$. These satisfy the following relations:
\be
N^{(k+1)}_{0,0}
 =  N^{(k)}_{1, 1}\,,
\label{eq:glue_rec2}
\ee
This is because when repeatedly multiplying the $\mbf{8}$'s, a singlet state
can only come from $\mbf{8}\times\mbf{8}$.
For $m > 0$ and $n > 0$,
 \be
 N^{(k+1)}_{m,0}
 & = &
 N^{(k)}_{m,0}
 + N^{(k)}_{m+1, 1}
 + N^{(k)}_{m-1, 2}
 + N^{(k)}_{m-2, 1}
 \\
 N^{(k+1)}_{0,n}
 & = &
 N^{(k)}_{0,n}
 + N^{(k)}_{2, n-1}
 + N^{(k)}_{1, n+1}
 + N^{(k)}_{1, n-2}
\label{eq:glue_rec3}
 \ee

 From \equ{eq:glue_rec1}, one can guess that the basic unit  in solving the
recursion
 relation should be
 \be
 C_{k:\; l_1, l_2, \cdots, l_6}
 \equiv
 {k!\over l_1!\, l_2!\, \cdots l_6!\, (k - \sum_{i=1}^6 l_i)!}
 2^{k-\sum_{i=1}^6 l_i} \, ,
 \ee
 which appears in the expansion of $8^k = (1 + 1 + 1 + 1 + 1 + 1 + 2)^k$.

Following the pattern established thus far in \equ{eq:glue_rec1},
we set
 \be
 (m, n)
 =
 l_1 (1, 1)
 +
 l_2 (-1, -1)
 +
 l_3 (2, -1)
 +
 l_4 (-2, 1)
 +
 l_5 (1, -2)
 +
 l_6 (-1, 2)
 \non
 \ee
One can then define the following function (again in analogy to section 4.1),
\be
 G_{k:\; m, n}
 =
 \sum_{l_1,\cdots, l_6=0}^k
 C_{k:\; l_1, l_2, \cdots, l_6}\,
 \delta_{m -[ (l_1 - l_2) + 2(l_3 - l_4) + (l_5 - l_6)]}\,
 \delta_{n -[ (l_1 - l_2) - (l_3 - l_4) - 2(l_5 - l_6)]}
 \non
\label{eq:gluesol1}
 \ee
The solution to
Eqs.(\ref{eq:glue_rec1}), (\ref{eq:glue_rec2}), and (\ref{eq:glue_rec3})
is again given by the same combination of $G_{k:\,m,n}$'s as in the quark case
 \be
 N_{m,n}^{(k)}
 & = &
 G_{k:\; m, n} + G_{k:\; m+3, n} + G_{k:\; m, n+3}
 \non
 & & {}
 -
 G_{k:\; m+2, n-1} - G_{k:\; m-1, n+2} - G_{k:\; m+2, n+2} \, .
 \label{eq:glue_sol}
 \ee
The validity of this form of solution has been checked extensively by
numerical means.  Namely, we generated $N_{m,n}^{(k)}$ numerically by using the
recursion relations and compared it with the values obtained by the
solution (\ref{eq:glue_sol}). We are therefore confident that
Eq.~(\ref{eq:glue_sol}) is the
solution.

To express $G_{k:\; m,n}$ in terms of the SU(3) Casimirs, we can use the
following
integral representation of the  Kronecker delta:
\be
\delta_{s,t} = \int_{-\pi}^\pi {d\phi\over 2\pi}\, e^{i\phi(s-t)}\, ,
\ee
which enables us to carry out the sum in Eq.~(\ref{eq:gluesol1}). This yields
\be
\lefteqn{G_{k:\; m, n}
=
\int_{-\pi}^\pi {d\phi\over 2\pi}\,
\int_{-\pi}^\pi {d\varphi\over 2\pi}\,
} &&
\non
& & {} \times
e^{i\phi m + i\varphi n}\,
\Bigg(2 +
2\cos(\phi+\varphi)+2\cos(2\phi-\varphi)
+2\cos(2\varphi-\phi)\Bigg)^k
\ee

Using the small angle approximation (valid for large $m, n$), one obtains,
\be
G_{k:\; m, n}& \approx &
8^k\,
{1\over 2\pi}\int_{-\infty}^\infty d\phi\,
{1\over 2\pi}\int_{-\infty}^\infty d\varphi\,
e^{i\phi m + i\varphi n}\,
\left(1
-(3/4)(\phi^2 + \varphi^2 - \phi\varphi)
\right)^k
\non
& \approx &
8^k\,
{1\over 2\pi}\int_{-\infty}^\infty d\phi\,
{1\over 2\pi}\int_{-\infty}^\infty d\varphi\,
e^{i\phi m + i\varphi n}\,
e^{-(3k/4) (\phi^2 + \varphi^2 - \phi\varphi)} \non
& = &
{8^k\over (2 \pi)^2} {8\pi\over 3 \sqrt{3} k}
e^{-4(m^2+n^2+mn)/(9 k)}\, .
\label{eq:glue_rec_soln}
\ee
Substituting this expression into Eq.~(\ref{eq:glue_sol})
 and applying the Taylor expansion technique, identically as in
 Eq.~(\ref{eq:approx_33}), we obtain
 for the multiplicity distribution of $k$ gluons
\be
N_{m,n}^{(k)}
&=& {8^k\over (2 \pi)^2}
  {\frac{512\,m\,n\,\left( m + n \right) \,\pi }
    {81\,{\sqrt{3}}\,
{k^4}}}\,\exp\left(-{C_F\over C_A}\, {N_c\over k}\, D_2^{m,n} \right)
\label{eq:glue_final}
 \ee

Eq.~(\ref{eq:glue_final})
is the final result of this sub-section.
It has several interesting features. Firstly, unlike the
case of many quarks, it has no dependence whatsoever on the cubic Casimir.
Indeed, the result is entirely
symmetric in $m$ and $n$. Secondly, the weight in the exponential is just
simply a Gaussian in the color charges.
These are of order $\sqrt{k} \gg 1$, and are therefore classical color charges.
The prefactor, as in the $SU(3)$ quark case, is proportional to $m\, n\,
(m+n)$.
Therefore, as in the $SU(3)$ case, one can write the sum over color charges in
a box as an integral over an 8-dimensional classical measure.
We further note that for the gluons, the total number of degrees of freedom
does turn
out to be $8^k$ even in our large $k$ approximation.
Finally,  the variance of the quadratic Casimir is
proportional to  an additional factor of $C_A/C_F$ compared to
the quark case. One
thus obtains for the
argument of the exponential (see Eq.~(\ref{eq:su2charge2})),
\be
{C_F\over C_A}\, {C_A\over k}\, {\bfQ}^2 &=& {C_F\over C_A}\, C_A\,
{(\Delta x_\perp)^4 \over g^2 k} \rho^a\rho^a\non
&=& {1\over 2\,N_c}\, {\pi R^2\over g^2 A}\, (\Delta x_\perp)^2 \rho^a \rho^a
\, ,
\ee
where, in our simple model $k$ is now the number of gluons in a tube of
transverse size $(\Delta x_\perp)^2$, namely,
$k= (N_c^2-1)\, A \, (\Delta x_\perp)^2 /\pi R^2$.
One can thus deduce that the
color charge squared per unit area
of the gluon charges is
\be
\mu_{A,{\rm glue}}^2 = g^2\, N_c \, {A\over \pi R^2} \, .
\ee
This result for the gluons is precisely the one expected in the MV
model~\cite{GyulassyMcLerran}.

The rest of the derivation of the path integral for the MV-effective action
goes through as discussed previously.
We end this discussion with a caveat. It has long been understood that a
renormalization group treatment
where one includes bremsstrahlung gluons from quark sources, as additional
sources for further bremsstrahlung
of softer ``wee" gluons, will modify the simple Gaussian of the MV
model~\cite{AJMV,Weigert,CGC}. Naively, one
can interpret this in terms of adding gluons to the color sources. The reason
we have a Gaussian distribution of sources in the problem discussed here is
two-fold; firstly, we ignore (quantum) correlations in the sources and
secondly, the ground state in our problem is a singlet ${\bf 1}$
representation. A
closer analogy to the bremsstrahlung scenario would be to have quarks in the
ground state and then add ${\bf 8}$
representations. This would correspond to different initial conditions for our
recursion relations and is outside the scope of this paper.

\subsection{Many $q{\bar q}$ pairs}

In this case, we are multiplying  $\bbox{3 \times \bar{3} = 1 + 8}$
representations.
Then, from the gluon case, it is straightforward to see that
 \be
 \lefteqn{\bbox{
 ((1,0) \times (0,1)) \times (m, n)}
 } &&
 \non
 & = &
 \bbox{
 (m, n) +
 (m+1, n+1) +
 (m+2, n-1)
 }
 \non
 & &
 \bbox{
 {} +
 (m+1, n-2) +
 (m-1, n+2)}
 \non
 & &
 \bbox{
 {} +
 (m, n) +
 (m, n) +
 (m-2, n+1) +
 (m-1, n-1)
 }
 \non
 \label{eq:qqbar}
 \ee
 The recursion relation for the multiplicity distribution $N_{m, n}$ is
identical
with the gluon
 case with the exception that one replaces $2N_{m,n}\rightarrow 3N_{m,n}$ in
 Eq.~(\ref{eq:gluons}). The
 special cases are also very similar to the gluon case. The basic unit in the
solution of Eq.~(\ref{eq:qqbar})
 are the ``n-nomial coefficients" in the expansion of the $9^k$ states as $9^k
= (1 + 1 + 1 + 1 + 1 + 1 + 3)^k$.

The rest of the derivation is very similar to the previous cases and we will
merely quote the result here. We
obtain in the large $k$ limit,
\be
N_{m,n}^{(k)} & \approx &
{9^k \over (2\pi)^2}\,
{\frac{27\,{\sqrt{3}}\,m\,n\,\left( m + n \right) \,\pi }
    {8\, {k^4}}}
{e^{-{\frac{{m^2} + m\,n + {n^2}}{2\,k}}}}
 \ee
Again, one finds an expression symmetric in $m$ and $n$, the  ubiquitous factor
$m\,n\,(m+n)$ in the prefactor, and
the Gaussian distribution of color charges. Since $D_2^{m,n} \approx (m^2 + n^2
+ mn)/3$, one can now
easily deduce that $\mu_{A,{\rm q\bar q}}^2 = 2\,\mu_A^2$.

\section{$SU(N_c)$}

We can carry out a similar analysis of the multiplicity of higher dimensional
representations for the
general $SU(N_c)$ case. For problems in the standard model,
it is sufficient to have explicit solutions for $SU(2)$ and $SU(3)$ cases.
However, there are several reasons to extend the analysis to the more
general $N_c$ case. Firstly, we are presented with an interesting non-trivial
random walk problem in $N-1$ dimensions  where the number of possible
directions at each step depends on the type of representation being
multiplied at that step.
Secondly, we would like to confirm that the patterns we see in the $N_c = 2$
and
$N_c=3$ cases are not accidental but follow from a more general solution.
For example, we interpreted the factors 2 and 3 that multiply the
quadratic Casimirs in Eqs.~(\ref{eq:su2prob}) and (\ref{eq:total_dof1}) as
the corresponding $N_c$ factors.
We would like to verify in this section that they
indeed are $N_c$ factors and not merely an accident of combinatorics.
Another pattern we observed in previous sections is that
the multiplicity is  predominantly determined
by the quadratic Casimir and that the most probable
distribution has weights of order $O(1/\sqrt{k})$. We also would like to see
that this is a general pattern.
Finally, the large $N_c$ limit constitutes another classical limit of a spin
system.
Although this large $N_c$ limit is not what one needs in the small $x$ MV
model, working out an explicit expression for a general $N_c$ may shed
some light on when exactly we can regard a system as a ``classical" system.

\subsection{$\mbf{N_c}$}

Let us consider adding a large number of quarks in the fundamental
representation of $SU(N_c)$.
For $SU(N_c)$, a representation $\mbf{R}$ can be uniquely
labeled by an integer vector
\be
\mbf{R} = (r_1, r_2, \cdots, r_{N_c-1}) \, .
\ee
The corresponding Young tableau has $N_c$ rows with $f_i$ number of boxes in
the $i$-th row. The number of boxes in the rows and $r_i$ are related by
$r_i = f_i - f_{i+1}$ and $f_i$'s should satisfy the inequality
\be
f_1 \ge f_2 \ge \cdots f_{N_c-1} \ge f_{N_c}
\ee
since $r_i \ge 0$.
In this notation, a single quark is labeled by
\be
\mbf{N_c} = (1, 0, \cdots, 0)
\ee

We begin with a random representation given by
\be
\mbf{R} = (r_1, r_2, \cdots, r_{N_c-1})
\ee
We would like to see what irreducible
representations are generated when $\mbfN_c$ and
$\mbfR$ are multiplied together.

Multiplying by $\mbfN_c$ corresponds to adding a box to the Young tableau
for $\mbfR$.
Adding a box to the $i$-th row lengthens it by 1 or $f_i \to f_i+1$.
Therefore the labels change as
\be
r_i = f_i - f_{i+1} \to r_i + 1
\\
r_{i-1} = f_{i-1} - f_{i} \to r_i - 1
\ee
The result of the multiplication by $N_c$ is
\be
\bf{\mbfR \times \mbfN_c}
& = &
\sum_{m=1}^{N_c} \mbf{R}'_m
\ee
where
\be
\mbfR'_m
& \equiv &
(r_1, r_2, \cdots, r_{N_c-1}) - \bfe_{m-1} + \bfe_{m}
\ee
Here
$\bfe_m$ is a unit vector in the $m$-th direction
(with $\bfe_0 = \bfe_{N_c} = 0$).
As before, $\mbfR'_m$ with negative entries should be discarded.
If one keeps multiplying $\mbf{N}_c$'s,
the multiplicities of the representations in the $k$-th iteration
and $(k{-1})$-th iteration are related by
\be
G^{(k)}(\mbfR)
&=&
\sum_{m=1}^{N_c}
G^{(k-1)}(\mbfR + \bfe_{m-1} - \bfe_m)
\ee
If $k \gg 1$ and $r_i \gg 1$, we can make the continuum approximation
by Taylor expanding the right hand side  and discarding higher derivative
terms:
\be
G^{(k)}(\mbfR)
\approx
N_c \,G^{(k-1)}(\mbfR)
+
\hatD_{N_c}
G^{(k-1)}(\mbfR)
\label{eq:pre_diff}
\ee
where we defined
\be
\hat{D}_{N_c} \equiv
\sum_{m=1}^{N_c-1}\partial_{m}^2
-
\sum_{m=1}^{N_c-2} \partial_{m}\partial_{m+1}\,,
\ee
with $\partial_s = {\partial/\partial r_s}$.

We also would like to make a similar
continuum approximation for $k$ as well so that Eq.~(\ref{eq:pre_diff})
becomes a partial differential equation.
Since $N_c$ can be large, care must be taken to ensure the validity of
the Taylor expansion.  To do so, we first let
\be
G^{(k)}(\mbfR) = G(k, \mbfR) = N_c^k \sigma(k, \mbfR) \,,
\ee
to take care of the potentially large $N_c$ dependence.
In the large $k$ limit, the equation for $\sigma(k, \mbfR)$ is then
\be
\partial_k \sigma = {1\over N_c}\hat{D}_{N_c}\sigma
\label{eq:su_n_eq}
\ee

Letting the expressions we obtained for the $SU(2)$ and $SU(3)$ cases guide us,
we try the following ansatz:
\be
\sigma(k,\mbfR) = g(k) \exp\left(-{D_2(\mbfR)\over rk}\right)\, ,
\label{eq:su_n_ansatz}
\ee
where $D_2(\mbfR)$ is the quadratic Casimir of the representation $\mbfR$
and $r$ is a yet to be determined constant.
Substituting Eq.~(\ref{eq:su_n_ansatz}) into Eq.~(\ref{eq:su_n_eq}) and
solving for $g(k)$ is tedious but straightforward.  The solution is
\be
G(k, \mbfR) = C\, {N_c^k\over k^{({N_c}-1)/2}}
\exp\left( -{N_c^2(N_c^2-1)\over 24 k} - N_c{D_2(\mbfR)\over k} \right) \, ,
\label{eq:su_N_sol}
\ee
where $C$ is a normalization constant.
Details of this derivation are in  Appendix C.

Note the appearance of $N_c D_2/k$ in the exponential.
Both the $SU(2)$ and $SU(3)$ cases we worked out match up with this
expression and the expression for the color charge squared per unit area
obtained in section 3.2
\be
\mu_A^2 = {g^2 A\over 2\pi R_A^2}\, ,
\ee
is valid for any $N_c$.
The additional term in the exponential only affects the overall
normalization and is negligible when $N_c \ll k$ as in the $SU(2)$ and
$SU(3)$ cases.

The expression in Eq.~(\ref{eq:su_N_sol}) is a particular solution of the
recursion relation in the large $k$ limit.  However, it is not guaranteed
that Eq.~(\ref{eq:su_N_sol}) satisfies the necessary boundary condition.
In fact, the solution (\ref{eq:su_N_sol}) corresponds to the
$G_{k:m,n}$
functions in the previous section.  The multiplicity $N(k, \mbfR)$ should
therefore be
a particular linear combination of $G(k, \mbfR)$ as in
Eq.~(\ref{eq:su3_q_sol}).

Unfortunately, we have been unable to find the analog of
Eq.~(\ref{eq:su3_q_sol}) for general $N_c$ although it is conceivable that a
general solution may be obtained along lines presented in Ref.\cite{Benkart}
and Ref.\cite{Tate}.
However, if the pattern seen in $SU(2)$ and $SU(3)$ were to persist,
one might guess that in the large $k$ limit, the multiplicity should be given
by
\be
N(k, \mbfR)
=
{\cal N}\,
d(\mbfR)\,
\exp\left(- N_c{D_2(\mbfR)\over k} \right)
\ee
where ${\cal N}$ is a normalization constant and $d(\mbfR)$ is the dimension of
the
$\mbfR$ representation in $SU(N_c)$.
We can also expect the color phase space measure will have the structure
\be
d^{N_c^2 - 1} Q
=
dr_1 dr_2\cdots dr_{N_c-1}\, d(\mbfR)^2\, d\Omega_{N_c\,(N_c-1)}
\ee
where $d\Omega_{N_c\,(N_c-1)}$ are the $\mbfR$ independent canonically
conjugate
Darboux variables. We have been unable thus far to verify these conjectures
explicitly.

\subsection{$\mbf{N}_c\times \bar{\mbf{N}}_c$}

We can easily generalize the above analysis to the case of a large number of
quark and anti-quark pairs.
For $SU(N_c)$, the fundamental and its dual representations are labeled by
\be
\mbf{N}_c = (1, 0, \cdots, 0)
\\
\bar{\mbf{N}}_c = (0, 0, \cdots, 1)
\ee

As previously, we start with a representation
\be
\mbfR = (r_1, r_2, \cdots, r_{N_c-1})
\ee
Multiplying by $\mbf{N}_c$ results in
\be
(r'_1, r'_2, \cdots, r'_{N_c-1}) =
(r_1, r_2, \cdots, r_{N_c-1}) - \bfe_{m-1} + \bfe_{m}
\ee
for $0 \le m \le N_c$.
Further multiplying by $\bar{\mbf{N}}_c$ results in
\be
(r''_1, r''_2, \cdots, r''_{N_c-1}) =
(r'_1, r'_2, \cdots, r'_{N_c-1}) + \bfe_{n-1} - \bfe_{n}
\ee
for $0 \le n \le N_c$.

Hence the multiplicities in the $k$-th iteration and the
($k{-}1$)-th iteration
are related by
\be
G^{(k)}(\mbfR)
=
\sum_{m,n=1}^{N_c}
G^{(k-1)}(\mbfR + \bfe_{m-1} - \bfe_m - \bfe_{n-1} + \bfe_n)
\label{eq:gkmbfr}
\ee
In the continuum limit, where all the $r_i$ are typically large (or $k$ is
large),
\be
G^{(k)}(\mbfR) =
N_c^2\, G^{(k-1)}(\mbfR)
+
2\,N_c\Bigg(
\sum_{m=1}^{N_c-1}\partial_{m}^2
-
\sum_{m=1}^{N_c-2} \partial_{m}\partial_{m+1}
\Bigg)
G^{(k-1)}(\mbfR)
\ee
We again let
\be
G^{(k-1)}(\mbfR)
=
G(k, \mbfR) = N_c^{2k}\, g(k)\, \exp\left(-{D_2(\mbfR)\over rk}\right)
\ee

The solution for the resulting differential equation is
\be
G(k, \mbfR)
=
C\,
{N_c^{2k} \over k^{(N-1)/2}}
\exp\left(
-{ N_c^2(N_c^2-1)\over 48 k}
-{N_c\, D_2(\mbfR)\over 2k}\right)
\ee
Again, this is a particular solution of the recursion relation.
The comments at the end of the last section (after Eq.~(\ref{eq:su_N_sol}))
are equally applicable here.

\subsection{The adjoint representation}

The adjoint case isn't much different from the quark-antiquark case.
For $SU(N_c)$,
\be
(\mbf{N_c^2-1}) = (1, 0, \cdots, 0, 1) \, .
\ee
We know that
\be
\mbf{N}_c\times \bar{\mbf{N}}_c = \mbf{1} \oplus (\mbf{N_c^2-1})
\ee
So all we have to do is to change
Eq.~(\ref{eq:gkmbfr}) to
\be
G^{(k)}(\mbfR)
=
\sum_{m,n=1}^{N_c}
G^{(k-1)}(\mbfR + \bfe_{m-1} - \bfe_m - \bfe_{n-1} + \bfe_n)
- G^{(k-1)}(\mbfR)
\ee
and the solution is
\be
G(k,\mbfR) =
{(N_c^2-1)^{k} \over k^{(N_c-1)/2}}
\exp\left(
-{ (N_c^2-1)^2\over 48 k}
-{(N_c^2-1)D_2(\mbfR)\over 2N_ck}\right)
\ee
Again, this is a particular solution of the recursion relation.
The comments after Eq.~(\ref{eq:su_N_sol}) also apply here.

\section{Discussion and Summary}

We have discussed in this paper the distribution of color charge
representations generated by $k$ partons in an $SU(N_c)$
gauge theory. Since the partons are random, the problem is a random walk
problem in the space spanned by the
$SU(N_c)$ Casimirs. We explicitly considered the $SU(2)$ and $SU(3)$ cases
before considering the general $N_c$ limit.
For all the cases considered, we find that the most likely representation is
one of order $O(\sqrt{k})$. The distribution of
representations about this representation is given by an exponential in the
quadratic Casimir $D_2$, with a weight proportional
to $k$. In the case of $SU(3)$ quarks, the contribution due to the cubic
Casimir is suppressed relative to the leading Gaussian term by $O(1/\sqrt{k})$.
Remarkably, for gluons and quark-anti-quark pairs, the result is given entirely
in terms of the quadratic
Casimir. Our results for $SU(3)$ can be generalized to the $SU(N_c)$ case.
Although the random walk is now in the space of
$N_c-1$ Casimirs, the distribution of representations is still given by the
quadratic Casimir.

Since the most likely representation is of order $O(\sqrt{k})$, one can argue
that, in the $\sqrt{k}\rightarrow \infty$ limit, the
representation is a classical representation. The arguments here are no
different from those for the treatment of classical
spins. A formal representation of
$SU(N_c)$ representations, suitable for path integral formulations,
can be made in terms of
coherent states~\cite{Mathur}.
As discussed in appendix A, it has been shown in Ref.~\cite{Gitman} that, for
large $k$, one recovers the classical limit.

Our discussion here was formulated in the framework of the
McLerran-Venugopalan model for small $x$ physics. The Gaussian
distribution of color charge sources and their treatment as classical color
charges was assumed in this model. We have provided
here a more rigorous basis for these assumptions. The small $x$ behavior of QCD
is more complex than the one outlined in
the MV model and more sophisticated renormalization group (RG) treatments have
been developed~\cite{JKLW,CGC}. A key feature of
the MV model, namely, the lack of correlations among the sources, breaks down
in these treatments. Therefore, unsurprisingly
in our random walk picture, the distributions are no longer Gaussian. The
classical assumption persists however since the most
probable representations are likely still classical ones. This assumption
deserves further study. Remarkably, in the limit of very small $x$,
a mean field treatment of the RG equations, recovers
a Gaussian distribution of sources-albeit a non-local one~\cite{IIM,Mueller}.
Interactions among sources in this limit produces screened charges of size
$1/Q_s$,  where $Q_s$ is
the saturation scale~\footnote{In the classical MV model, $Q_s^2\approx
\mu_A^2$, where $\mu_A^2$ is the color charge squared per unit area.}. In our
language, this might mean that the effective charges are random and can
therefore be represented by
Gaussian sources. It is interesting to speculate whether the combinatorial
techniques developed here for higher
dimensional representations can be used in numerical studies of high energy
Onium-Onium scattering in QCD~\cite{IM}.

Other clear applications of the techniques developed here are in transport
problems in QCD. It has been argued previously
that the transport of color in high temperature QCD could be represented in
terms of the classical dynamics of color charges~\cite{Kelly}. It was however
not clear when this classical description was the appropriate one and attempts
to derive these from first principles have had to resort to ad hoc
approximations~\cite{JSRJ}. We do know though that any
systematic treatment of transport problems involves coarse graining of color
charges over large distance scales.  These
can therefore be treated using the recursion techniques developed here. These
developments will be discussed elsewhere~\cite{JVW}.

\section*{Acknowledgments}

We would like to thank Volker Koch, Alex Kovner, 
Xin-Nian Wang and especially J\o rgen Randrup
for early discussions on the topics discussed 
in this paper. We thank Jean-Paul Blaizot for re-awakening our interest in 
classical representations in small $x$ QCD. Thanks go to Dmitri Diakonov and Chris Korthals Altes for useful group 
theory discussions. Finally, we would like to thank Keijo Kajantie 
for reading the manuscript. 
SJ would like to acknowledge the support of the
Nuclear Science Division in LBL where this work was initiated. 
He is supported in part by the Natural Sciences and
Engineering Research Council of Canada and by le Fonds 
Nature et Technologies of Qu\'ebec.  
SJ also
thanks RIKEN BNL Center and U.S. Department of Energy [DE-AC02-98CH10886]
for providing facilities essential for the completion of this work.
RV thanks the Physics Dept. of McGill Univ.
for their hospitality. His research is
supported by DOE Contract No. DE-AC02-98CH10886.

\appendix
\section{Large $k$ color charge representations as classical color charge
representations}

An interesting limit of quantum theories is the limit of $N\rightarrow \infty$,
where $N$ can denote either the underlying invariance group of the theory or
the size of higher dimensional representations when the invariance group of the
theory is held fixed.
In both cases, the large $N$ limit corresponds to a classical limit and can be
formally shown to correspond to the limit
where $\hbar \rightarrow 0$. As discussed in a nice review by
Yaffe~\cite{Yaffe}, the quantum dynamics of a system will
reduce to classical dynamics as $\hbar \rightarrow 0$, if and only if the
system can be prepared in a state whose uncertainty
in its conjugate momenta and positions vanishes in this limit.

Coherent states~\cite{Klauder} have this property and it is therefore useful to
write states of the quantum theory in this representation. These coherent
states are orbits of the Heisenberg-Weyl group, and can therefore be
generalized to
any Lie group. For an $SU(N_c)$ gauge theory, one can define coherent states
carrying $SU(N_c)$ color charges~\cite{Mathur}.

In Ref.~\cite{Gitman}, coherent states for $SU(N_c)$ groups were studied
with the particular purpose of studying
the classical limit. The authors identify coherent states of a particular
representation in terms of the signature of representation
(which is none other than size of the totally symmetric tensor-in our case $k$~\cite{KorthalsAltes})
and the weight vectors ${\bf n}$ of the
representation. They define the variance of the square of the length of the
isospin vector (the quadratic Casimir $D_2 = \sum Q_a^2$),
\be
\Delta D_2 = \langle\Psi| \sum_a Q_a^2|\Psi\rangle
- \sum_a \langle \Psi|Q_a|\Psi\rangle^2 \equiv
\langle \Psi|D_2 - D_2^\prime|\Psi \rangle \, ,
\ee
In the coherent state basis for  fixed $SU(N_c)$,
\be
{\Delta D_2\over D_2} = {N_c\over N_c +k} \,.
\ee
This result is valid for $k \gg N_c$.
Thus the large $k$ limit, where the variance vanishes can be identified with
the $\hbar \rightarrow 0$ limit in quantum mechanics,
and one can formally relate $\hbar =1/k$. We refer interested readers to
Ref.~\cite{Gitman} for a more extensive discussion
of this correspondence.

\section{The volume of the canonical phase space in SU(3)}

In Refs.\cite{Johnson,Kelly},
the phase space variables for $SU(3)$ are parameterized
as
\be
 Q_1 &=& \cos\phi_1\pi_+ \pi_-, \ \ \   Q_2 = \sin\phi_1 \pi_+ \pi_-
\non
 Q_3 &=& \pi_1
\non
 Q_4 &=& C_{++}\pi_+ A + C_{+-}\pi_- B,\ \ \ Q_5 = S_{++}\pi_+ A + S_{+-}\pi_-
B
\non
 Q_6 &=& C_{-+}\pi_-A - C_{--}\pi_+ B,\ \ \
 Q_7 = S_{-+} \pi_- A- S_{--} \pi_+ B
\non
 Q_8 &=& \pi_2
\ee
with
\be
\pi_+ & = & \sqrt{\pi_3 + \pi_1}, \ \ \ \pi_- = \sqrt{\pi_3 - \pi_1}
\non
C_{\pm\pm}
&=& \cos\left({1\over 2}(\pm\phi_1 + \sqrt{3}\phi_2 \pm \phi_3)\right),
\ \ \
S_{\pm\pm}
= \sin\left({1\over 2}(\pm\phi_1 + \sqrt{3}\phi_2 \pm \phi_3)\right),
\non
\ee
and
\be
A &=&
{1\over 2\pi_3}
\sqrt{
\left({J_1-J_2\over 3}+\pi_3+{\pi_2\over \sqrt{3}}\right)
\left({J_1+2J_2\over 3}+\pi_3+{\pi_2\over \sqrt{3}}\right)
\left({2J_1+J_2\over 3}-\pi_3-{\pi_2\over \sqrt{3}}\right)
}
\non
B &=&
{1\over 2\pi_3}
\sqrt{
\left({J_2-J_1\over 3}+\pi_3-{\pi_2\over \sqrt{3}}\right)
\left({J_1+2J_2\over 3}-\pi_3+{\pi_2\over \sqrt{3}}\right)
\left({2J_1+J_2\over 3}+\pi_3-{\pi_2\over \sqrt{3}}\right)
}
\non
\ee
where $(J_1, J_2)$ equals the $(m, n)$ label of a $SU(3)$
representation.  The pairs $(\pi_i, \phi_i)$ form the canonical pairs.
The total phase space volume given for a fixed $(m, n)$ is given by the
integration over these canonical variables:
\be
\Omega(m,n) = \int_{m,n} d\phi_1 d\pi_1 d\phi_2 d\pi_2 d\phi_3 d\pi_3
\label{eq:Omega}
\ee

Evaluating the integral over the
conjugate momenta is not trivial.
It can  be inferred from Fig.~1 of Johnson's paper~\cite{Johnson} but the
explicit
derivation, as shown below, is quite involved (if straightforward).

The values of the canonical momenta $\pi_i$ are restricted to make $A$ and
$B$ real.
 To solve for the ranges of $\pi$'s, it is convenient to
 define
 \be
 & K_1 \equiv {2m + n\over 3}, \ \ \ K_2 \equiv {2n + m\over 3}  &
 \non
 &x =  \pi_3 + {\pi_2\over \sqrt{3}}, \ \ \
 y  =  \pi_3 - {\pi_2\over \sqrt{3}} &
 \ee
 which also gives
 \be
 dx dy =
 d(\pi_3 + \pi_2/\sqrt{3})
 d(\pi_3 - \pi_2/\sqrt{3})
 =
 {2\over \sqrt{3}} d\pi_3 d\pi_2
 \ee
 The factors in $A$ and $B$ are
\be
A_1 &=& K_1 - K_2 + x
\\
A_2 &=& K_2 + x
\\
A_3 &=& K_1 - x
\\
B_1 &=& K_2 - K_1 + y
\\
B_2 &=& K_2 - y
\\
B_3 &=& K_1 + y
\ee
 Potentially, there are 16 sign combinations for the factors $A_i$ and $B_i$
 that make the products $A_1 A_2 A_3$ and $B_1 B_2 B_3$ positive.
 For instance, to make the argument of $A$ positive,
 one may require all the factors in $A$ to be positive or require that
 $A_1$ and $A_2$ to be negative and $A_3$ to be positive.

 Fortunately, due to the conditions $x + y> 0$, $K_1 > 0$, $K_2 > 0$ and
 \be
 2K_1 - K_2 = m \ge 0
 \\
 2K_2 - K_1 = n \ge 0
 \ee
one can easily show that the only consistent sign combination is
for all $A_i$ and $B_i$ to be positive.

 The condition that all $A_i$ are positive yields the range
 \be
 K_2-K_1 < x < K_1
 \ee
The condition that all $B_i$ are positive yields the range
 \be
 K_1 - K_2 < y < K_2
 \ee
 then
 \be
 \int_{K_2-K_1}^{K_1} dx\, \int_{K_1-K_2}^{K_2} dy\,
 \int_{-(x+y)/2}^{(x+y)/2} d\pi_1
 & = &
 \int_{K_2-K_1}^{K_1} dx\, \int_{K_1-K_2}^{K_2} dy\, (x+y)
 \non
 & = &
 {mn(m+n)\over 2}
 \ee
Equivalently, 
 \be
 \int d\pi_1 d\pi_2 d\pi_3 = {\sqrt{3}\over 4} m n (m+n)
\label{eq:pi_int}
 \ee
 
 The integrals over the angles can be evaluated in the following
 way. We first note that
\be
\int d^8 Q \delta(Q^2 - 1)
& = &
\int dQ\, Q^7\, \int d\Omega_7 \delta(Q^2 - 1)
\non
& = &
{\pi^4\over 6}\,,
\ee
where we have used the fact that the area of a unit sphere in 8-D is ${2\pi^4/3!}$~\footnote{For a nice 
discussion of the volume and area of $SU(N)$ groups, we refer the readers to ~\cite{group_vol}.}.
On the other hand, using Eq.(\ref{eq:darboux1}) this can be also written as
\be
{\pi^4\over 6}
& = &
\int d^8Q \delta(Q^2 - 1)
\non
& = &
\int d\phi_1 d\phi_2 d\phi_3 d\pi_1 d\pi_2 d\pi_3 dm dn\,
mn(m+n){\sqrt{3}\over 48} \delta\left({m^2+mn+n^2\over 3} - 1\right)
\non
\ee
We already know the result of the $\pi_i$ integration from Eq.~(\ref{eq:pi_int}). Therefore, 
\be
{\pi^4\over 6}
= 
{1\over 64}\int d\phi_1 d\phi_2 d\phi_3 
dm dn\, 
m^2n^2(m+n)^2
\delta\left({m^2+mn+n^2\over 3} - 1\right)
\ee
The delta function gives
\be
m_1 = {\frac{-n + {\sqrt{3}}\,{\sqrt{4 - {n^2}}}}{2}}
\ee
and the condition $m \ge 0$ becomes $n \le \sqrt{3}$.
Hence
\be
\int dm dn\,
m^2n^2(m+n)^2 \delta\left({m^2+mn+n^2\over 3} - 1\right)
& = &
\left. \int_0^{\sqrt{3}} dn\, {3(m^2 n^2 (m+n)^2)\over 2m + n}
\right|_{m=m_1}
\non
& =&
\int_0^{\sqrt{3}} dn\, 
{\frac{{\sqrt{3}}\,{n^2}\,
{{\left( -3 + {n^2} \right) }^2}} {{\sqrt{4 - {n^2}}}}} 
\non
& = &
{2\pi\over \sqrt{3}} \, ,
\ee
which results in 
\be
\int d\phi_1 d\phi_2 d\phi_3 
& = &
{2 (2\pi)^3\over \sqrt{3}} \, .
\label{eq:phi_int}
\ee
The total phase space volume defined in Eq.~(\ref{eq:Omega}) is therefore
determined from Eq.~(\ref{eq:pi_int}) and Eq.~(\ref{eq:phi_int}) to be  
\be
\Omega(m,n) = {(2\pi)^3\over 2}m n (m+n) \, .
\ee

\section{Young tableaux and the quadratic Casimir in $SU(N_c)$}

 In this section, we establish the relationship between a Young tableau and the
 quadratic Casimir used in the text to obtain $SU(N_c)$ results.

 For $SU(N_c)$, a Young tableau has $N_c$ rows.
 Suppose a Young tableau has $f_j$ boxes in the $j$-th row.
 Then this tableau corresponds to a representation labeled by
 \be
 \mbfR = (r_1, r_2, \cdots, r_{{N_c}-1})
 \ee
 where $r_i = f_i - f_{i+1}$.
 Inverting this relationship yields
 \be
 f_i = \sum_{j=i}^{N_c} r_j\,,
 \ee
 with
 \be
 r_{N_c} = {1\over N_c}\left(K -\sum_{j=1}^{N_c-1} j r_j \right)\,,
 \ee
 where
 \be
 K = \sum_{i=1}^{N_c} f_i
 \ee
 is the total number of boxes in the tableau.

 According to Okubo~\cite{Okubo}, the quadratic Casimir for $SU(N_c)$ is given
by
 \be
 D_2(\mbfR)
 = {C_2(\mbfR)\over 2}
 \ee
 where
 \be
 C_2(\mbfR)
 & = &
 K(1 + N_c - K/N_c)
 +
 \sum_{j=1}^{N_c} f_j(f_j - 2j)
\label{eq:okubo_D2}
 \ee
 Since $D_2$ is a Casimir that depends only on $\mbfR$, it must be
 independent of $K$ even if the expression (\ref{eq:okubo_D2}) explicitly
 contains $K$. The fact that $C_2$ is in fact independent of $K$ can be
 easily checked by taking a derivative with respect to $K$ and using that
 fact that only $r_N$ depends on $K$.

 In the main text, the equation we need to solve has the form
 \be
 \partial_k \sigma(k, \mbfR) = {a\over N_c} \hatD_{N_c}\sigma(k, \mbfR)
 \label{eq:dksigma}
 \ee
 where
\be
\hat{D}_{N_c} \equiv
\sum_{m=1}^{N_c-1}\partial_{m}^2
-
\sum_{m=1}^{N_c-2} \partial_{m}\partial_{m+1}
\ee
Our ansatz is
\be
\sigma(k, \mbfR) = g(k) \exp\left(-{C_2(\mbfR)\over rk}\right)
\ee
The left hand side is then
\be
\partial_k \sigma =
\partial_k g \, e^{-C_2/rk}
+ {C_2\over rt^2} g\, e^{-C_2/rk}
\ee
where we used the fact that $C_2$ is independent of $K$ or equivalently, $k$.
The right hand side is
\be
\hat{D}_{N_c} g\, e^{-C_2/rk}
& = &
g \hat{D}_{N_c} e^{-C_2/rk}
\non
& = &
{1\over r^2 k^2}
\left(
\sum_{m=1}^{N_c-1}(\partial_{m} C_2)^2
-
\sum_{m=1}^{N_c-2} (\partial_{m} C_2) (\partial_{m+1} C_2)
\right) g e^{-C_2/rk}
\non
& & {}
-
{1\over rk}
\left(
\sum_{m=1}^{N_c-1}\partial_{m}^2 C_2
-
\sum_{m=1}^{N_c-2} \partial_{m}\partial_{m+1} C_2
\right) g e^{-C_2/rk}
\ee
 Since the expression Eq.~(\ref{eq:okubo_D2}) depends on $f_j$,
 we need to know
 \be
 \partial_m f_l
 & = &
 {\partial\over \partial r_m}
 \left(
 r_l + r_{l+1} + \cdots + r_{N_c}
 \right)
 \non
 & = &
 (1-m/{N_c}) - \theta(l-m)
 \ee
 where
 $\theta(l-m)=1$ if $l > m$ and $\theta(l-m) = 0$ if $l \le m$.
 Therefore,
 \be
 \partial_m C_2
 & = &
 2\sum_{l=1}^{N_c} f_l \partial_m f_l
 -
 2\sum_{l=1}^{N_c} l \partial_m f_l
 \non
 & = &
 2\sum_{l=1}^{N_c} f_l
 \left(
 (1-m/{N_c}) - \theta(l-m)
 \right)
 -
 2\sum_{l=1}^{N_c} l
 \left(
 (1-m/{N_c}) - \theta(l-m)
 \right)
 \non
 & = &
 -2(m/{N_c}) K
 +
 m({N_c}-m)
 +
 2\sum_{l=1}^m f_l
 \ee
 where we used the fact that $\sum_{l=1}^{N_c} f_l = K$.
 We note that this formula works even if $m = {N_c}$ or $m=0$.
 In that case $\partial_0 C_2 = 0$,
 $\partial_{N_c} C_2 = 0$.

Second derivatives are
 \be
 \partial_m^2  C_2
 & = &
 \partial_m 2 \sum_{l=1}^m f_l
 =
 2 \partial_m (K- \sum_{l=m+1}^{N_c} f_l )
 \non
 & = &
 -2 \sum_{l=m+1}^{N_c} \left(-{m\over {N_c}}\right)
 \non
 & = &
 2 {m({N_c}-m)\over {N_c}}
 \ee
and
 \be
 \partial_{m+1} \partial_m  C_2
 & = &
 -2
 \partial_{m+1}
 \sum_{l=m+1}^{N_c} f_l
 \non
 & = &
 2 {m({N_c}-m-1)\over {N_c}}
 \ee

Using these expressions, it is tedious but straightforward to work out
\be
 \sum_{m=1}^{{N_c}-1} \partial_m^2 C_2
 -
 \sum_{m=1}^{{N_c}-2} \partial_{m+1} \partial_m C_2
 & = &
 ({N_c}-1)
 \ee
 and
 \be
 \sum_{m=1}^{{N_c}-1} (\partial_m C_2)^2
 -
 \sum_{m=1}^{{N_c}-2} (\partial_{m+1} C_2)( \partial_m C_2)
 & = &
 2\, C_2 + {{N_c}({N_c}^2-1)\over 6}
 \ee
The equation  (\ref{eq:dksigma}) then becomes
\be
\partial_k g
+ {C_2\over rk^2} g
& = &
{a\,g\over {N_c}}
\Bigg[
{1\over r^2 k^2}
\left(
2C_2 + {{N_c}({N_c}^2-1)\over 6}
\right)
-
{1\over rk}
\left(
{N_c}-1
\right)
\Bigg] 
\ee
We first let
\be
r = {2a\over {N_c}}
\ee
then solve
\be
\partial_k \ln g
& =&
{{N_c}^2({N_c}^2-1)\over 24 a k^2}
-
{1\over 2k}({N_c}-1)
\ee
which yields
\be
\sigma(k,\mbfR) = {S\over k^{({N_c}-1)/2}} \exp\left(
-{{N_c}^2({N_c}^2-1)\over 24 a k} - {N_c}{C_2(\mbfR)\over 2 a k}
\right)
\ee
where $S$ is an arbitrary normalization constant.

\end{document}